\documentclass{elsarticle}

\usepackage{graphicx}
\usepackage{amsmath}
\usepackage{hyperref}
\usepackage{dcolumn}

\newcommand{\Al}{Al(BH$_4$)$_3$}

\journal{International Journal of Hydrogen Energy}

\begin{document}

\begin{frontmatter}

\title{Suppressing diborane production during the hydrogen release
of metal borohydrides: The example of alloyed \Al}

\author{D. Harrison}

\author{T. Thonhauser\corref{mycorrespondingauthor}}
\cortext[mycorrespondingauthor]{Corresponding author}
\ead{thonhauser@wfu.edu}
\address{Department of Physics, Wake Forest University,
Winston-Salem, NC 27109, USA.}

\date{\today}

\begin{abstract}
Aluminum borohydride (\Al) is an example of a promising hydrogen storage
material with exceptional hydrogen densities by weight and volume and a
low hydrogen desorption temperature. But, unfortunately, its production
of diborane (B$_2$H$_6$) gases upon heating to release the hydrogen
restricts its practical use. To elucidate this issue, we investigate
the properties of a number of metal borohydrides with the same problem
and find that the electronegativity of the metal cation is not the best
descriptor of diborane production. We show that, instead, the closely
related formation enthalpy is a better descriptor and we find that
diborane production is an exponential function thereof. We conclude
that diborane production is sufficiently suppressed for formation
enthalpies of $-$80~kJ/mol~BH$_4$ or lower, providing specific design
guidelines to tune existing metal borohydrides or synthesize new ones.
We then use first-principles methods to study the effects of Sc alloying
in \Al. Our results for the thermodynamic properties of the
Al$_{1-x}$Sc$_x$(BH$_4$)$_3$ alloy clearly show the stabilizing effect
of Sc alloying and thus the suppression of diborane production. We
conclude that stabilizing \Al\ and similar borohydrides via alloying or
other means is a promising route to suppress diborane production and
thus develop viable hydrogen storage materials.
\end{abstract}




\end{frontmatter}

\newpage

\section{Introduction}\label{sec:introduction}

Hydrogen has been identified as a promising alternative to fossil fuels
due to its potential to be a clean, renewable energy
carrier \cite{Kunowsky_2013:material_demands,
Durbin_2013:review_hydrogen, Crabtree_2004:hydrogen_economy}. However, in
its natural state, hydrogen has an unsuitably low volumetric density for
automotive applications, leading to much research on more effective
hydrogen storage methods \cite{Harrison_2015:materials_hydrogen,Ahluwalia_2012:on-board_off-board}, with
clear goals outlined by the Department on Energy (DOE)
\cite{Yang_2010:high_capacity, DOE_Targets_Onboard_2009}. Borohydrides
are a class of complex hydrides which are of interest due to their high
gravimetric and volumetric densities, although they typically suffer
from hydrogen desorption temperatures over 85~$^\circ$C, i.e.\ the
maximum delivery temperature set by the DOE for fuel cell operation in
vehicles \cite{Yang_2010:high_capacity, Graetz_2009:new_approaches,
Ronnebro_2011:development_group, Li_2011:recent_progress,
Rude_2011:tailoring_properties,
Jain_2010:novel_hydrogen,Orimo_2007:complex_hydrides,Umegaki_2009:boron-_nitrogen-based,George_2010:structural_stability}.

Of particular interest is aluminum borohydride (\Al), due in part to its
high storage density of 16.9~mass\%, but primarily because of its
significantly lower initial hydrogen release temperature of 334~K
(61~$^\circ$C) compared to other
borohydrides \cite{Li_2008:dehydriding_rehydriding,
Nakamori_2007:thermodynamical_stabilities}. Despite these advantages,
\Al\ has not received as much attention as other borohydrides---partly
due to its cost, but mainly because of diborane production during its
hydrogen release \cite{Dalebrook_2013:hydrogen_storage,
Lodziana_2010:multivalent_metal}. In addition to reducing H$_2$ output,
diborane is also poisonous to the fuel cell, reducing fuel cell
performance by 39\%\ while present, although full recovery is retainable
once diborane is removed from the fuel stream \cite{V_B_4_2011}. This
production of diborane is due to the instability of the material itself:
the difference in energy between the \Al\ molecule and solid phases is
rather small, causing solid \Al\ to be composed of discrete molecular
\Al\ units held together by weak van der Waals interactions. These weak
interactions among the \Al\ molecules are also responsible for its low
melting point of 209~K \cite{Miwa_2007:first-principles_study,
Lodziana_2010:multivalent_metal}.

Many methods have been used in order to alter the properties of various
borohydrides, including destabilizing via reactions with other hydrides
\cite{Vajo_2005:reversible_storage, Alapati_2007:using_first,
Alapati_2008:large-scale_screening, Li_2008:dehydriding_rehydriding},
alloying \cite{Nickels_2008:tuning_decomposition,Li_2007:materials_designing,Lee_2010:effect_mg}, cation substitution
\cite{Setten_2007:ab_initio}, anion substitution
\cite{Brinks_2008:adjustment_stability}, and adding catalysts
\cite{Li_2007:effects_ball}. In general, the goal of most of these
methods is to lower the hydrogen desorption temperature by altering
either the kinetics or thermodynamics of the reaction. Here, we proposed
to use the same kind of methods, but instead take a material with an
already low desorption temperature and try to suppress its undesirable
diborane production.

\begin{figure}[t]
\centering\includegraphics[width=0.75\columnwidth]{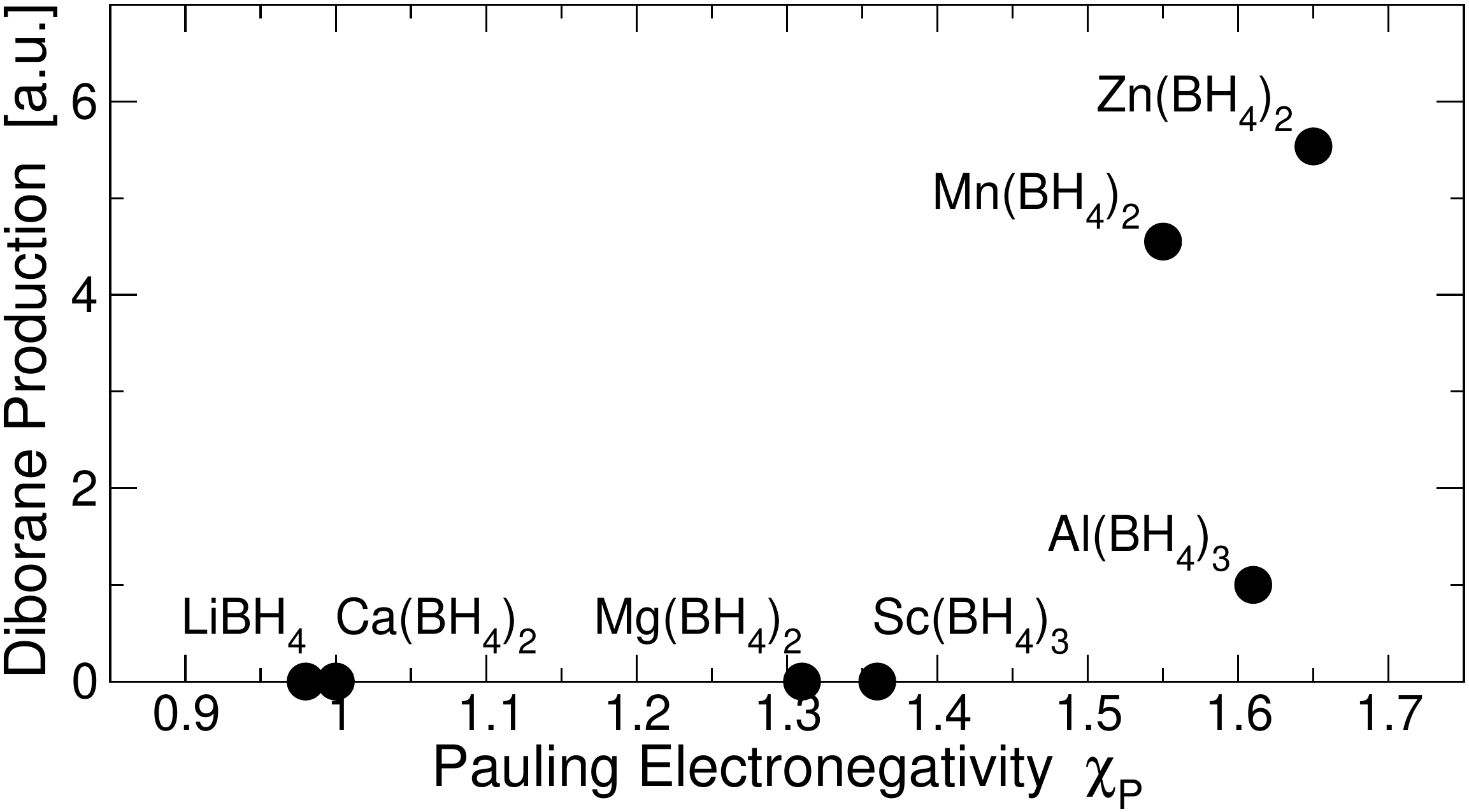}
\caption{\label{fig:pauling}Plot of diborane production vs.\ $\chi_P$
for various metal borohydrides. Diborane production data was obtained by
integrating mass-spectroscopy data from Nakamori et
al.~\cite{Nakamori_2007:thermodynamical_stabilities}. Values for $\chi_P$
for the metal cation are taken from
Ref.~\cite{Allred_1961:electronegativity_values}. Amount of diborane
produced does not follow a clear trend as a function of $\chi_P$.}
\end{figure}

The production of diborane in a borohydride is correlated to the Pauling
electronegativity $\chi_P$ of its metal cation, as established by
Nakamori et al.\ \cite{Nakamori_2006:correlation_between,
Nakamori_2007:thermodynamical_stabilities}. In Fig.~\ref{fig:pauling}, we
show $\chi_P$ vs.\ the amount of diborane production (relative to
hydrogen produced) as measured by integrating the mass-spectroscopy
data \cite{Nakamori_2007:thermodynamical_stabilities}; for a given
borohydride, we took the ratio of the integrated B$_2$H$_6$ data to
the integrated H$_2$ data, normalized so that \Al\ has a value of 1.
As can be seen, the amount of diborane produced does not follow any
clear trend as a function of $\chi_P$. The only thing that can be said
is that somewhere between $1.36<\chi_P<1.55$ diborane production starts.
These are rather course boundaries, providing little guidance for tuning
borohydrides, and they do not reveal information about the amount of
diborane produced. A better predictor is thus desirable.

It is also known that, as the stability of a borohydride is largely due
to the charge transfer between the metal cation and the anionic
[BH$_4$]$^-$ unit, $\chi_P$ is also strongly correlated with the
enthalpy of formation \cite{Miwa_2007:first-principles_study}. We
hypothesized that by lowering the enthalpy of formation of \Al\ (i.e.\
stabilizing it), the production of diborane upon heating is suppressed.
Because the enthalpy of formation of borohydrides is so strongly
correlated with the cation electronegativity, alloying with a more
(less) electronegative cation is an effective method to increase
(decrease) the formation enthalpy. As seen in our previous work and
other works studying the effect of alloying on borohydrides, we expect
that the enthalpy of formation of the alloyed material is approximately the weighted
average of the two materials in their pure state (i.e.\ the enthalpy of
mixing is small) \cite{Harrison_2014:tuning_hydrogen,
Albanese_2013:theoretical_experimental}.

In choosing which metal to alloy \Al\ with, we try to satisfy three
criteria: (i) the metal should have a lower electronegativity than
aluminum, (ii) it should have a valence of 3, and (iii) it should be
relatively lightweight in order to preserve the high hydrogen mass
density for the alloyed material. For these reasons, we chose scandium,
which in addition to fulfilling these criteria is known to form a stable
borohydride Sc(BH$_4$)$_3$ that does not produce diborane gas upon
decomposition \cite{Nakamori_2007:thermodynamical_stabilities}. For
reference, a number of studies on other borohydride alloys involving Sc
or Al have already been reported \cite{Cerny_2010:nascbh44_novel,
Cerny_2010:structure_characterization, Dovgaliuk_2014:first_halide-free,
Kim_2009:liscbh44_hydrogen, Lindemann_2013:insight_into,
Yang_2007:destabilizing_libh4}.

In order to demonstrate the effect of scandium on \Al, we first verify
that Sc alloying stabilizes \Al\ by calculating how the
temperature-dependent enthalpy of formation changes with Sc
concentration. However, in order to estimate the amount of Sc needed to
alloy \Al, $\chi_P$ cannot be used as a predictor for the reasons
explained above. In the search for a better predictor, we determine the
relationship between the stability of the borohydride (i.e.\ formation
enthalpy) and the amount of diborane produced. To this end, we consult
the mass-spectroscopy data of Nakamori et al., wherein we analyze their
data in order to determine the relative diborane production of \Al,
Zn(BH$_4$)$_2$, and Mn(BH$_4$)$_2$, the only borohydrides to produce
diborane in any significant
amount \cite{Nakamori_2007:thermodynamical_stabilities}; we then
calculate the thermodynamics of these borohydrides so that we can find a
more quantitative relationship between formation enthalpy and diborane
production. Using this relationship as well as data from previous
borohydride alloy studies, we estimate the ideal diborane production.

\section{Computational Details}

\subsection{General}
\label{sec:general}

In order to obtain high-accuracy results on the thermodynamics of \Al, we
performed {\it ab initio} simulations using density-functional theory (DFT) as implemented in \textsc{Vasp}
\cite{Kresse_1996:efficient_iterative, Kresse_1999:ultrasoft_pseudopotentials}.
We used the standard projector augmented wave (PAW) pseudopotentials included in \textsc{Vasp} with a
450~eV kinetic energy cutoff. The energy convergence criterion was
$10^{-7}$~eV. The structure used for \Al\ was the solid-phase $\beta$-\Al\
taken from the theoretical work of Miwa et al.\
\cite{Miwa_2007:first-principles_study,Aldridge_1997:some_tetrahydroborate},
the structure for Mn(BH$_4$)$_2$ was taken from the theoretical work of Choudhury et al.\ \cite{Choudhury_2009:manganese_borohydride}, the structures
for Sc(BH$_4$)$_3$ and Zn(BH$_4$)$_2$ were taken from Nakamori et al.\ \cite{Nakamori_2006:correlation_between},
and the structure for B was the 106 atom $\beta$-rhombohedral structure
suggested by van Setten et al.~\cite{Setten_2007:thermodynamic_stability}. A
fixed $10\times10\times10$~\AA\ unit cell was used for gas-phase H$_2$. The $k$-point meshes for \Al\ (including Sc alloying),
Mn(BH$_4$)$_2$, Zn(BH$_4$)$_2$, B, Al, Sc, Mn, and Zn were $2\times2\times1$,
$2\times2\times2$, $2\times3\times2$, $2\times2\times2$, $15\times15\times15$,
$21\times21\times12$, $5\times5\times5$, and $17\times17\times7$, respectively.
All k-point meshes were converged within 1~meV per atom with respect to much
larger meshes.  All
structures were relaxed with respect to unit-cell parameters and atom positions
until all forces were less than 0.2~meV/\AA. 

Sc-alloying was done by replacing a random Al atom with a Sc atom and then relaxing
the structure; as the unit cell of \Al\ contains 4 Al atoms, 25, 50, and 75\% alloying
was done by replacing 1, 2, and 3 Al atoms with Sc. Energy differences due to which atom was replaced were found to be small
(less than 3~kJ/mol) compared to the energy difference due to alloying. Frequency calculations were done
with the symmetry-reduced finite-displacement method with the recommended
displacement of 0.015~\AA. Supercells used in the frequency calculations for
\Al, Mn(BH$_4$)$_2$, Zn(BH$_4$)$_2$, B, Al, Sc, Mn, and Zn were
$2\times2\times1$, $1\times1\times1$, $2\times2\times2$, $1\times1\times1$,
$3\times3\times3$, $5\times5\times3$, $1\times1\times1$, and $5\times5\times3$,
respectively.

Previous studies on \Al\ found that van der Waals interactions are
important to accurately model the material due to weak interactions
among the discrete molecular \Al\
units \cite{Miwa_2007:first-principles_study,
Lodziana_2010:multivalent_metal}. For this reason, in order to obtain
accurate thermodynamic results, we employed the exchange-correlation
functional vdW-DF \cite{Thonhauser_2015:spin_signature,
Berland_2015:van_waals, Thonhauser_2007:van_waals,
Langreth_2009:density_functional}, which includes a truly nonlocal
correlation term to capture van der Waals binding. In order to ascertain
the effect of van der Waals interactions, comparisons to previous
studies are made below.

\subsection{Enthalpies, Entropies, and Mixing}

The temperature dependent vibrational contribution to the enthalpy and
entropy is calculated as
\begin{eqnarray}
H_{\rm vib}&=&\int_0^\infty d\omega
   \bigg(\frac{1}{2}+\frac{1}{\exp[\hbar\omega/kT]-1}\bigg)
   g(\omega)\hbar\omega\;,\label{eq:enthalpy_phonon}\\[1ex]
S_{\rm vib} &=& \int_0^\infty d\omega
   \bigg(\frac{\hbar\omega}{2T}\coth\frac{\hbar\omega}{2kT}
   -k\ln\Big[2\sinh\frac{\hbar\omega}{2kT}\Big]\bigg)
   g(\omega)\;,\mspace{30mu}\label{eq:entropy_phonon}
\end{eqnarray}
where $T$ is the temperature, $k$ is Boltzmann's
constant, $\omega$ is the vibrational frequency, and $g(\omega)$ is the phonon
density of states. The phonon density of states was calculated using the program
phonopy \cite{Togo_2008:first-principles_calculations}. The total enthalpy
then is the sum of the DFT ground-state energy and this vibrational
contribution. Note that the zero-point energy (ZPE) is already included
in Eq.~(\ref{eq:enthalpy_phonon}). The formation enthalpy is
calculated as the difference in the enthalpies of the material and its
constituent elements (thus, elements in their natural state have
formation enthalpies of 0). We then can calculate the enthalpies and
entropies of reaction, using the differences in formation enthalpies and absolute entropies
of all materials. One notable exception is H$_2$: due to it being a gas
we calculated its ground state and zero-point energy, but took the
temperature-dependent contribution from experiment, following the
approach of van Setten et al.\
\cite{Setten_2008:density_functional,Hemmes_1986:thermodynamic_properties}.
Particularly, we used
$H_{\rm{H_2\,gas}}(T)=E_{\rm{H_2}}+E^{\rm{ZPE}}_{\rm{H_2}}+H^{\rm{exp}}_{\rm{H_2\,gas}}(T)$,
where the electronic energy $E_{\rm{H_2}}$ and zero-point energy
$E^{\rm{ZPE}}_{\rm{H_2}}$ were calculated using DFT and the last term
was taken from known experimental values \cite{Hemmes_1986:thermodynamic_properties}.
For reactions involving Sc alloying, the entropy of mixing was
calculated according to
\begin{equation}
S_{\rm mix}= -k_B\; [c\ln{c}+(1-c)\ln{(1-c)}]\;,
\end{equation}
where $c$ is the concentration of Sc, but it was found to be negligible.

\section{Results and Discussion}

\subsection{Structure of \Al\ and Importance of van der Waals Interactions}

As mentioned, van der Waals interactions were found to be an important
factor in accurately modeling \Al\ due to the weak interactions among
the discrete molecular units \cite{Miwa_2007:first-principles_study,
Lodziana_2010:multivalent_metal}. A previous study on \Al\ by Miwa et
al.\ found lattice constants for $\beta$-\Al\ of $a=18.649$~\AA,
$b=6.488$~\AA, and $c=6.389$~\AA, whereas we find values of
$a=17.917$~\AA, $b=6.026$~\AA, and $c=6.022$~\AA, in much better
agreement with the experimental values of $a=18.021$~\AA, $b=6.138$~\AA,
and $c=6.199$~\AA\
\cite{Miwa_2007:first-principles_study,Aldridge_1997:some_tetrahydroborate}.
As described by Miwa et al., this discrepancy is likely due to weak van
der Waals interactions among molecular \Al\ units, which standard
functionals are unable to describe. Miwa et al.\ also calculated the
energy of a single molecule of \Al\ and found an energy difference with
respect to the solid of only $\sim$10~kJ/mol, while we, using vdW-DF,
calculate a much higher energy difference of 64.6~kJ/mol. To verify our
results, we also did an energy calculation on both the solid and
molecule without vdW-DF and find an energy difference of only
5.2~kJ/mol---in good agreement with the value from Miwa et al. Clearly
then, the nonlocal van der Waals interactions are responsible for the
majority of the stability of \Al, and including them is crucial to
obtain the most accurate treatment of this material.

\subsection{Structure of Alloyed \Al}

As described in Section \ref{sec:general}, 
Sc-alloying was done by replacing a random Al atom in the \Al\ structure
with a Sc atom and then
relaxing the system. It is conceivable that for higher Sc concentrations
the true ground-state structure of the alloys prefer structures that cannot
be reached that way, but a truly-random structure search for the alloys is
beyond the scope of this paper. However, it is interesting that by analyzing
our alloy structures, we observe the trend that the substituted Sc
pulls nearby BH$_4$ units close to it. Indeed, we would expect
in general for the BH$_4$ units to cluster around the less electronegative cation,
as more stable borohydrides (and thus possessing
a less electronegative metal) are also more dense and have a higher
coordination of BH$_4$ units near the metal \cite{George_2010:structural_stability}.

\subsection{Formation Enthalpies}

Table \ref{tab:formation_enthalpies} shows the formation enthalpies of
\Al\ for various values of Sc-alloying. The formation enthalpy linearly
decreases with increasing Sc concentration, showing that Sc alloying
predictably stabilizes \Al\ and that the thermodynamic properties of the
alloy are nearly the weighted average of the properties of the pure
borohydrides, i.e.\ \Al\ and Sc(BH$_4$)$_3$. 

\begin{table}
\caption{\label{tab:formation_enthalpies}Formation enthalpies $\Delta
H_f$ in kJ/mol; the first column is using only the DFT ground-state
energy, the second includes the zero-point energy (ZPE) correction, and
the third includes the lattice vibration contribution to the energy at
300~K.}
\begin{tabular*}{\columnwidth}{@{}l@{\extracolsep{\fill}}rrr@{}}\hline\hline
Material & $\Delta H_f^{E_{\rm only}}$ & $\Delta H_f^{\rm ZPE}$ & $\Delta H_f^{\rm 300K}$ \\\hline
\Al\                                &  $-168.12$  &    $-52.67$      & $ -83.88$ \\
Al$_{3/4}$Sc$_{1/4}$(BH$_4$)$_3$    &  $-214.06$  &    $-100.03$     & $-131.92$ \\
Al$_{1/2}$Sc$_{1/2}$(BH$_4$)$_3$    &  $-259.39$  &    $-146.60$     & $-179.22$ \\
Al$_{1/4}$Sc$_{3/4}$(BH$_3$)$_3$    &  $-318.64$  &    $-206.14$     & $-240.04$ \\
Sc(BH$_4$)$_3$                      &  $-339.85$  &    $-229.49$     & $-265.12$ \\\hline\hline
\end{tabular*}
\end{table}

When we compare our $E_\text{only}$ formation enthalpy from
Table~\ref{tab:formation_enthalpies} of \Al\ calculated with vdW-DF to
the results of Miwa et al.~\footnote{Here, we are comparing our
formation enthalpy without ZPE, i.e.\ $E_\text{only}$, as this is what
was calculated in Ref.~\citenum{Miwa_2007:first-principles_study}.}, we
find a surprising disparity: our value of $-$168.12~kJ/mol to their
value of $-$131~kJ/mol \cite{Miwa_2007:first-principles_study}. Because
of this, we also calculated the formation enthalpy of \Al\ without
including van der Waals interactions with vdW-DF, and found a value of
$-$130~kJ/mol, in excellent agreement with Miwa et al. This significant
difference shows again that the inclusion of van der Waals interactions
is particularly important to achieving highly accurate results for \Al,
even more-so than for other borohydrides.

\subsection{Reaction Enthalpies at 300 K and 1 Bar}

\begin{table*}
\caption{\label{tab:react_values}Reaction enthalpies $\Delta H_r$ in
kJ/mol~H$_2$ and entropies $\Delta S_r$ in J/K/mol~H$_2$ at 300 K for
several \Al\ desorption reactions. The critical temperature $T_c$
predicted from the van't Hoff equation $\ln{p}=-\Delta H/RT+\Delta S/R$
for 1 bar H$_2$ pressure is given in K.}
\begin{tabular*}{\textwidth}{@{}lr@{\quad$\longrightarrow$\quad}l@{\extracolsep{\fill}}rrr@{}}\hline\hline
No.& Reactants                         & Products & $\Delta H_r^{\rm 300K}$ & $\Delta S_r^{\rm 300K}$ & $T_c$ \\\hline
1  & \Al\                              & $\textrm{Al}+3\textrm{B}+6\textrm{H}_2$                                    & 13.98  & 107.26  & 130 \\
2  & Al$_{3/4}$Sc$_{1/4}$(BH$_4$)$_3$  & $\frac{3}{4}\textrm{Al}+\frac{1}{4}\textrm{Sc}+3\textrm{B}+6\textrm{H}_2$  & 21.99  & 107.49  & 205 \\
3  & Al$_{1/2}$Sc$_{1/2}$(BH$_4$)$_3$  & $\frac{1}{2}\textrm{Al}+\frac{1}{2}\textrm{Sc}+3\textrm{B}+6\textrm{H}_2$  & 29.87  & 108.57  & 275 \\
4  & Al$_{1/4}$Sc$_{3/4}$(BH$_3$)$_3$  & $\frac{1}{4}\textrm{Al}+\frac{3}{4}\textrm{Sc}+3\textrm{B}+6\textrm{H}_2$  & 40.01  & 110.36  & 363 \\
5  & Sc(BH$_4$)$_3$                    & $\textrm{Sc}+3\textrm{B}+6\textrm{H}_2$                                    & 44.19  & 113.87  & 388 \\\hline\hline 
\end{tabular*}
\end{table*}

Table \ref{tab:react_values} shows the reaction enthalpies for the
decomposition reactions of \Al\ for various level of Sc-alloying at
300~K and 1~bar. For consistency's sake, we consider here the
decomposition into the elements in order to estimate the stabilizing
effect of Sc-alloying and to compare to other theoretical works---of
course, the true decomposition reaction for \Al\ is known to produce
diborane gas as well as other aluminoborane intermediates, as discussed
at length by \L{}odziana \cite{Lodziana_2010:multivalent_metal}.

From looking at the critical temperature for Reaction~1 of 130~K,
although the reaction shows \Al\ to be unstable with respect to its
elements at room temperature, it is in fact kinetically stabilized, due
to the previously mentioned diborane and aluminoborane intermediates of
the true hydrogen desorption reaction. Comparing our value from
Reaction~1 of 13.98~kJ/mol~H$_2$ to the value found by \L{}odziana of
11.7~kJ/mol~H$_2$ (taken from Fig.~1 of
Ref.~\cite{Lodziana_2010:multivalent_metal}), we again find our value
predicting greater stability for \Al. This confirms that vdW-DF
stabilizes \Al\ on the order of 15 -- 25\%\ more than standard
generalized gradient approximation (GGA) functionals.

From Table \ref{tab:react_values} we see that the critical temperature
for the hydrogen release of the alloy is roughly the weighted average of the
critical temperature of the pure borohydrides, i.e.\ based on our
results we would expect approximately a 70~K increase per 25\% Sc
alloying. However, as the hydrogen release in borohydrides is well known
to have a large kinetic barrier, the true relationship is likely more
complex. In fact, a study by Paskevicius et al.\ found that borohydride
alloys exhibit both lower melting temperatures and lower hydrogen
release temperatures than either of the pure borohydrides constituting
the alloy \cite{Paskevicius_2013:eutectic_melting}. In light of this, it
is difficult---and beyond the scope of this paper---to accurately
estimate the hydrogen release temperature of the alloy, other than to
say we expect the full release reaction to have a minimum temperature as
dictated by its thermodynamics.

\subsection{Relationship between Formation Enthalpy and Diborane Production}

\begin{figure}
\centering\includegraphics[width=0.75\columnwidth]{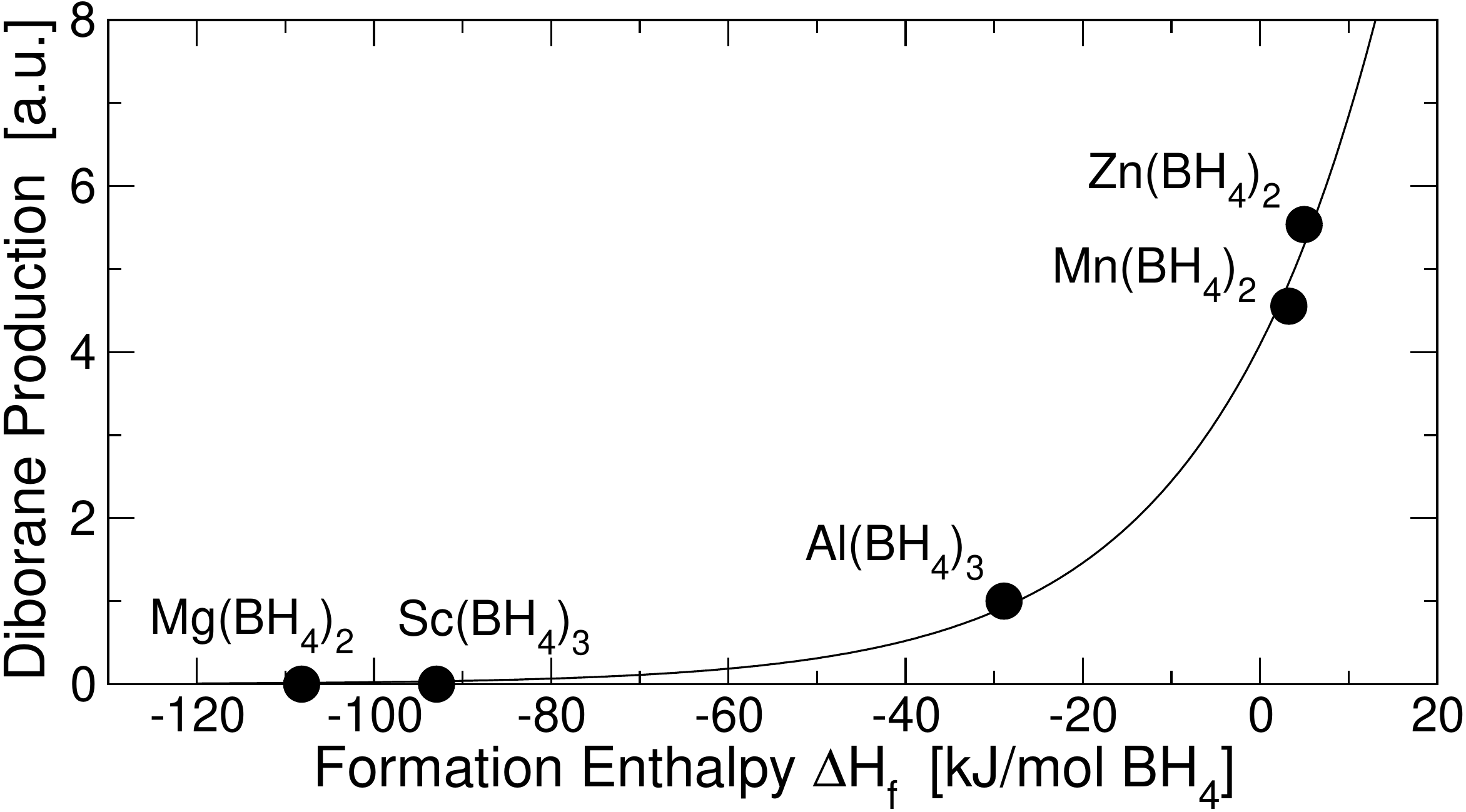}
\caption{\label{fig:diborane_production}Plot of diborane production
versus calculated formation enthalpy [kJ/mol~BH$_4$]. Diborane
production data was obtained by integrating mass-spectroscopy data from
Nakamori et al.~\cite{Nakamori_2007:thermodynamical_stabilities}. The
amount of diborane produced follows nicely an exponential behavior as a
function of formation enthalpy, as indicated by the fitted line
$\alpha\exp[\,\beta \Delta H_f]$ with $\alpha=4.089$ and
$\beta=0.05148$.}
\end{figure}

In order to estimate the concentration of Sc needed to suitably suppress
diborane production, we plot in Fig.~\ref{fig:diborane_production} the
amount of diborane produced (relative to hydrogen produced) versus
formation enthalpy for \Al, Zn(BH$_4$)$_2$, and Mn(BH$_4$)$_2$. The data
for diborane production was obtained by integrating the
mass-spectroscopy data from Nakamori et al.\ (units are arbitrary); for
a given borohydride, we took the ratio of the integrated B$_2$H$_6$ data
divided by the integrated H$_2$ data, normalized so that \Al\ has a
value of 1 \cite{Nakamori_2007:thermodynamical_stabilities}. The
formation enthalpy was calculated by us using DFT, with the formation
enthalpy taken at the approximate temperature of the full desorption in
the thermogravimetry curve, i.e.\ 340~K, 400~K, and 470~K for \Al,
Zn(BH$_4$)$_2$, and Mn(BH$_4$)$_2$, respectively.

As mentioned above, the Pauling electronegativity is only a coarse
predictor of a borohydride's tendency to produce diborane. From the data
in Fig.~\ref{fig:diborane_production} it is apparent that formation
enthalpy---and thus borohydride stability---is a more accurate
predictor. Whereas the Pauling electronegativity would predict \Al\ to
produce more diborane than Mn(BH$_4$)$_2$, due to the higher
electronegativity of Al compared to Mn (1.61 compared to 1.55)
\cite{Allred_1961:electronegativity_values}, the formation enthalpy
correctly predicts what actually happens: the less stable Mn(BH$_4$)$_2$
produces significantly more diborane than \Al.

The data in Fig.~\ref{fig:diborane_production} can be nicely fitted to
an exponential function as $\alpha\exp[\,\beta \Delta H_f]$ with
$\alpha=4.089$ and $\beta=0.05148$, although $\alpha$ is really
arbitrary as we have chosen to normalize the value for \Al\ to 1. An
exponential fit---suggested by the Arrhenius equation---can be justified
by first assuming that the formation enthalpy of a borohydride is
proportional to the reaction enthalpy for diborane formation (this is
reasonable, as replacing B and H$_2$ by B$_2$H$_6$ simply shifts all of
the formation enthalpies by a constant), and then assuming that the
reaction enthalpy for diborane formation is proportional to the
activation enthalpy. It follows that, in order to reduce diborane
production in \Al\ down to less than 5\%\ of its current value, i.e.\ 1
to 0.05, we need an enthalpy of formation of $-85.5$~kJ/mol~BH$_4$.

Unfortunately, as there is experimental data only for 3 borohydrides which produce diborane in any
significant amount, we cannot fully rely on this exponential fit.
Furthermore, it is worth mentioning that---while we assume here the
experimental data is for the pure borohydrides, and find this to be the
most likely case---there have been known
difficulties synthesizing certain borohydrides; for example,
attempts to synthesize Sc(BH$_4$)$_3$ have been known to form
LiSc(BH$_4$)$_4$ under
certain synthesis conditions \cite{Hagemann_2008:liscbh44_novel}.
For this reason, we also look to other studies on borohydride alloys which
have successfully suppressed diborane production in order to further
verify our criteria for diborane production. 
In particular, we found two
experimental studies involving Mg--Zn borohydride alloys done by the
same groups
\cite{Albanese_2013:theoretical_experimental,Kalantzopoulos_2014:hydrogen_storage}.
It was found that diborane production in a Mg$_{1-x}$Zn$_x$(BH$_4$)$_2$
alloy was dependent on the concentration of Zn; for values of $x$
greater than 0.248 (i.e.\ 24.8\%~Zn), at least small amounts of diborane
were measured while for values of $x$ less than 0.167 no diborane was
found. Comparing this to the known formation enthalpies for
Mg$_{1-x}$Zn$_x$(BH$_4$)$_2$ alloys (taken at 500~K)
\cite{Harrison_2014:tuning_hydrogen}, we find the cutoff for diborane
formation in the Mg--Zn alloy between $-88$ and $-79$~kJ/mol~BH$_4$, in
excellent agreement to our predicted value of $-85.5$~kJ/mol~BH$_4$ for
\Al. Combining the exponential fit of thermodynamic data of known
borohydrides with this data from Mg$_{1-x}$Zn$_x$(BH$_4$)$_2$ alloys, we
thus estimate the cutoff formation enthalpy for diborane production in
general to be around $-80$~kJ/mol~BH$_4$. This estimate for the formation
enthalpy includes both the zero-point energy and temperature effects.
Adjusting it to be suitable for energy-only calculations---which are
significantly simpler than including zero-point motion and temperature
effects---our data suggests a cutoff of $-110$~kJ/mol~BH$_4$.

We note that this estimate for the formation enthalpy required to
suppress diborane applies to either pure borohydrides or solid solutions of
different borohydrides, as this is the data we have included in our model.
However, our finding might even be more general, as the following example
suggests. A study by Gu et al.\ found that diborane
release is fully suppressed for Zn(BH$_4$)$_2$ when it is combined with
ammonia to form
Zn(BH$_4$)$_2\cdot2$NH$_3$~\cite{Gu_2012:structure_decomposition}. A recent 2015
study on ammine metal borohydrides claims that ammonia, when combined with
borohydrides which have a highly electronegative cation (as in the case with
Zn(BH$_4$)$_2$), stabilizes the
borohydride~\cite{Jepsen_2015:tailoring_properties}. Doing a simple energy
calculation for the formation enthalpy of Zn(BH$_4$)$_2\cdot2$NH$_3$ (i.e., not
including zero-point energy or temperature effects), we find a value of
$-165$~kJ/mol~BH$_4$, well below the energy-only threshold mentioned above
of $-110$~kJ/mol~BH$_4$. For reference, the energy-only formation enthalpy of Zn(BH$_4$)$_2$ is only
$-17$~kJ/mol~BH$_4$, while for Sc(BH$_4$)$_3$, which we
know does not produce diborane, it is $-113$~kJ/mol~BH$_4$. From this we can see
that NH$_3$ significantly stabilizes Zn(BH$_4$)$_2$ and this stabilization is
likely responsible for the suppression of diborane. Our criteria
for diborane suppression seems to apply not only to borohydride alloys, but
also to borohydrides modified through other means, such as additives. It is
thus conceivable that our guidelines for diborane suppression are applicable
to an even larger group of materials. If this is the case, our criteria could
be used in conjunction with a variety of methods in order to design borohydrides
which both have a low desorption temperature and no diborane production.
Examples already exist where such a criteria could be useful, e.g.\ a recent study
found that Mg(BH$_4$)$_2$ exhibits a significantly
reduced hydrogen desorption temperature when combined with fluorographite \cite{Zhang_2015:remarkable_enhancement}, however,
this hydrogen release is accompanied by diborane gas. Even more interesting, when LiBH$_4$ is added,
so that a Mg(BH$_4$)$_2$/LiBH$_4$/fluorographite composite is formed, a
low hydrogen desorption temperature combined with no diborane gas formation is observed.
It is very likely that adding the LiBH$_4$ stabilized the structure, suppressing the diborane formation.

Applying our estimate for the cutoff formation enthalpy to the case of
Sc doping in \Al, we find that approximately 80\% Sc concentration is
required to suppress diborane production. It is worth noting at this
point that, due to the cost of Sc, it is not a suitable hydrogen storage
material in any large amount; certainly not as 80\% of an alloy. But, we
see the strength and novelty of our study in that we were able to find a
precise predictor of diborane production in metal borohydrides and
provide accurate guidelines for the tuning of already existing
borohydrides and the design of new ones. Note that the formation
enthalpy can also be tuned by other means, such as stress, which is
already work in progress. We had chosen Sc due to it being one of the
most stable borohydrides with the same valence as \Al\, making it an
ideal alloyant for a case study. Furthermore, other studies on
borohydride alloys have found that certain metal combinations do not
seem to suppress diborane production (e.g.\ Li--Al and Li--Mn alloys),
most likely due to the phase separation of the solid solution, causing
the constituents to undergo their decomposition reactions separately
\cite{Lindemann_2013:insight_into,Paskevicius_2013:eutectic_melting}.
For example, Li et al.\ found that an alloy formed with LiBH$_4$ and Zn(BH$_4$)$_2$ or Al(BH$_4$)$_3$ disproportionate, possibly due to the
incompatible structures between the borohydrides \cite{Li_2007:materials_designing}.
It is important that the individual borohydrides constituting the alloy
have similar structures, as in the
case of a Al--Sc and Mg--Zn borohydride alloy, so as to prevent phase separation.

\section{Conclusions}

\emph{Ab initio} calculations were performed to study the hydrogen
storage material \Al\ and the effect of Sc alloying. It was found that
the thermodynamic properties of Sc-alloyed \Al\ are approximately the weighted average
of the properties of pure \Al\ and Sc(BH$_4$)$_3$. It was furthermore
found that proper accounting for van der Waals interactions is necessary
to achieve an accurate formation enthalpy due to the weak interactions
among individual \Al\ molecules in the \Al\ solid. In attempting to
estimate the ideal concentration of Sc needed to suppress diborane
production, we have found the formation enthalpy of a borohydride to be
a better predictor than the Pauling electronegativity, and estimate
approximately 80~at\%~Sc would be needed to suppress diborane
production. However, work done on diborane suppression in borohydrides
is relatively scarce and we encourage further work on Al--Sc borohydride
alloys at multiple concentrations. We have shown that, alternatively to
trying to \emph{destabilize} candidate hydrogen storage materials such
as LiBH$_4$ with too high of a hydrogen release temperature via
alloying, it is also possible to \emph{stabilize} materials such as \Al\
with an already low hydrogen desorption temperature and thus suppress
the unwanted diborane byproduct. Because of \Al's especially low
hydrogen release temperature compared to other borohydrides, modifying
\Al\ primarily so as to not produce diborane, especially via alloying
with more stable borohydrides, may be a promising avenue for developing
a viable hydrogen storage material within the class of borohydrides.


\section*{References}


\end{document}